\documentclass[preprint]{elsarticle}
\usepackage{latexsym, graphicx}

\begin{document}

\begin{frontmatter}

\title{Instantaneous spatially local projective measurements are consistent in a relativistic quantum field}
\author[syl]{Shih-Yuin Lin}
\ead{sylin@cc.ncue.edu.tw}
\address[syl]{Department of Physics, National Changhua University of Education,
Changhua 50007, Taiwan}
\date{August 17, 2012}

\begin{abstract}
Suppose the postulate of measurement in quantum mechanics can be extended to quantum field theory, then a local projective
measurement at some moment on an object locally coupled with a relativistic quantum field will result in a projection or collapse
of the wavefunctional of the combined system defined on the whole time-slice associated with the very moment of the measurement, 
if the relevant degrees of freedom have nonzero correlations. This implies that the wavefunctionals in the same Hamiltonian 
system but defined in different reference frames would collapse on different time-slices passing through the same local event 
where the measurement was done. Are these post-measurement states consistent with each other? 
We illustrate that the quantum states of the Raine-Sciama-Grove detector-field system started with the same initial Gaussian 
state defined on the same initial time-slice, then collapsed by the measurements on the pointlike detectors on different 
time-slices in different frames, will evolve to the same state of the combined system up to a coordinate transformation when 
compared on the same final time-slice.
Such consistency is guaranteed by the spatial locality of interactions and the general covariance in a relativistic system, 
together with the spatial locality of measurements and the linearity of quantum dynamics in its quantum theory.
\end{abstract}






\end{frontmatter}

\section{Introduction}

Quantum nonlocality in quantum mechanics (QM) manifests when combining quantum entanglement of two or more parties 
with quantum measurement on one of these parties. In the simplest scenario a quantum measurement \emph{locally} 
(in a subspace of the full Hilbert space) on one party of an entangled pair (say, $A$ and $B$) will lead to an 
instantaneous projection or collapse of the quantum state of both parties such that the other party is also affected.

The situation becomes more intriguing if we add the assumption that both $A$ and $B$ are local in position space 
and they are separated at a distance. Then the quantum state of $B$ will be projected instantaneously by a  
local measurement on $A$, no matter how far $B$ is away from $A$ \cite{EPR35}.
While this appears to have some kind of superluminal signal, causality will not be violated since no
meaningful information can be communicated using such an instantaneous wavefunction collapse. 
Aharonov and Albert have further shown that indirect measurement on quantum objects localized in space and time
is consistent with relativistic QM \cite{AA81}, where 
the measurement process has no covariant description in terms of the time evolution of quantum states \cite{AA84} 
and quantum states make sense only within a given frame. 

Nevertheless, local quantum objects such as atoms or charged particles are inevitably coupled with quantum 
fields defined on the whole spacelike hypersurface (the time-slice) and evolving in time as the environment. 
If a relativistic quantum field is involved in the system that we are looking at, will the above scenario 
of wavefunction collapse still be consistent? Or more generally, is instantaneous wavefunction collapse in 
a projective measurement local in position space consistent with relativistic quantum field theory (RQFT)?

RQFT is very powerful in solving the scattering problem. In particle colliders, measurements on an ensemble of 
particles are done in the huge detectors surrounding the collision point, effectively in the future asymptotic region. 
So it is sufficient to calculate the scattering amplitude between the in- and 
out-states defined in past infinity and future infinity, respectively, where all particles are free. This is called the 
``in-out" formalism, which gives the statistics of the outgoing particles against the incoming particles \cite{Hat92}. 
Our questions raised here, however, concern the {\it single-shot} projective measurements on a quantum field at some moment 
\emph{in the interaction region} rather than in the asymptotic region. So we have to go beyond the ``in-out" formalism to answer 
our questions.

One is tempted to generalize the indirect measurement scheme 
in \cite{AA81} to RQFT to study this issue. Nevertheless, there each indirect measurement process is modeled by an interaction 
localized in space and time between a quantum probe and the fields, and the projective measurements on the quantum probes 
are still performed in the future asymptotic region (more discussion will be given in Appendix A), while 
the wavefunction collapse we are looking at is not described by any interaction Hamiltonian.
Thus we turn to the standard Schr\"odinger picture of RQFT to watch the discontinuous and continuous evolutions of the whole system 
\cite{Hat92}, whose quantum state at each moment is represented as the wavefunctional of the fields (and the sources, if any) living on the 
whole associated time-slice. In this formalism one needs to specify an initial state on the time-slice associated with some non-infinity 
initial moment in some coordinate where the Hamiltonian is defined (we assume that it is always possible to prepare such an initial state), 
then the Schr\"odinger equation will give the continuous evolution of the quantum state from the initial moment and between the events
of the projective measurement.

Suppose the postulate of the projective measurement in QM can be extended to RQFT, then the wavefunctional of the fields on 
the whole time-slice would be collapsed by a measurement local in position space if the relevant degrees of freedom have nonzero correlations.    
Since the dynamical variables of quantum fields can be nonlocal in position space or separable in a quantum state, 
such a scenario of local measurement must be carefully formulated. 
One simple way to achieve this is by going back to the atom-field interacting system: to measure an Unruh-DeWitt detector 
\cite{Unr76, DeW79, LH06, LH07} or similar object locally coupled with quantum fields \cite{AA81, AA84, BP02, Dio91}, 
analogous to an optical system with a photodiode coupled to EM field, as is described in our setup.

Below we are looking at, but not limited to, a detector model similar to the Unruh-DeWitt detector theory.
Before getting into detailed calculation, in Section II we give an alternative frame in Minkowski space 
to make our discussion more economic and precise. Then in Section III we introduce the Raine-Sciama-Grove detector theory in 
(1+1)D Minkowski space as our toy model. We perform explicit calculations for one-detector and two-detector cases in 
Sections IV and V, respectively, then the results will be summarized in Section VI. Finally, we compare our model 
with those in the indirect measurement scheme in Appendix A, 
and apply an argument similar to that for our linear toy model to nonlinear detector-field models in Appendix B.

\section{An alternative frame in Minkowski space}

It has been shown by Aharonov and Albert that, in relativistic QM, quantum states {\it in the non-asymptotic region} defined 
on two different time-slices intersecting at some spacetime points could be very different even  
for the sectors of the dynamical variables defined right on the intersections of the two time-slices \cite{AA84}. 
Thus in the interaction region one can compare two wavefunctionals of a field defined in two different frames 
only if the whole time-slices that they are living on are exactly the same.

\begin{figure}
\includegraphics[width=5cm]{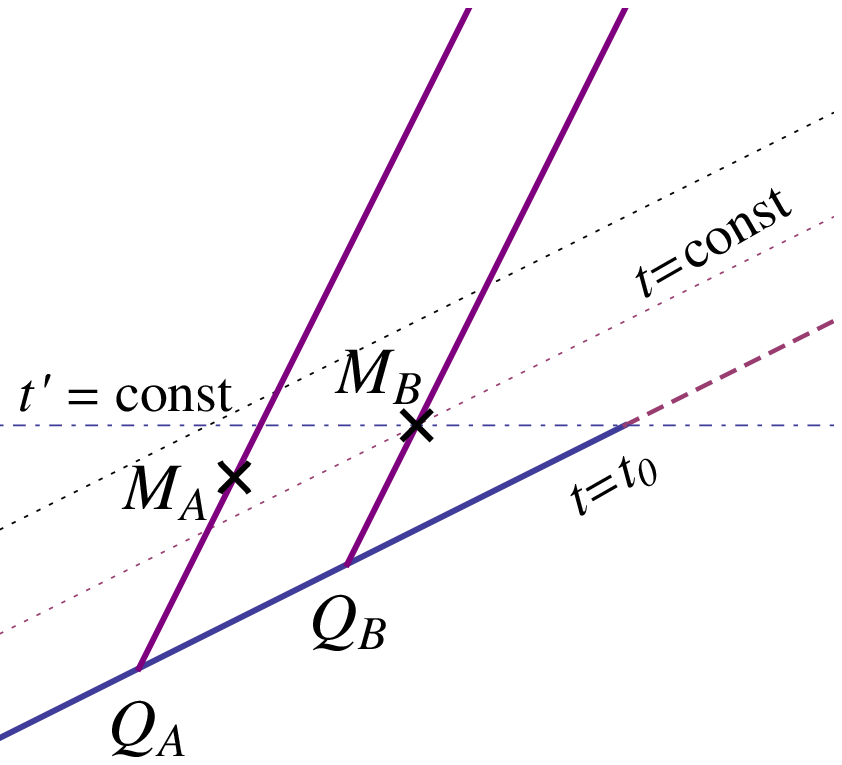}
\includegraphics[width=5cm]{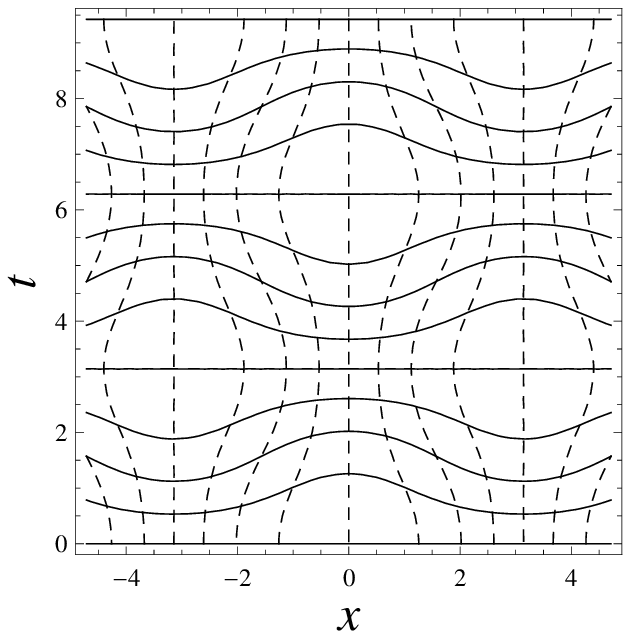}
\caption{(Left) Time order of two spacelike separated measurement events $M_A$ and $M_B$ on
$Q_A$ and $Q_B$, respectively, can
be altered by a Lorentz transformation: $M_B$ is earlier than $M_A$ in $t$, but later in $t'$. 
Here the dotted lines denote the $t$-slices, while the dot-dashed line is a constant $t'$ hypersurface. 
Note that the initial data of the quantum field on the dashed-line part 
of the $t_0$-slice appear to be specified {\it after} $M_A$ and $M_B$ for an inertial observer in 
coordinate time $t'$. 
(Right) The alternative coordinates $(\eta, \xi)$ given by $(\ref{etadef})$ with $A=1/2$ 
in the $t$-$x$ diagram of (1+1)D Minkowski space. The solid and dashed curves are constant $\eta$ 
and $\xi$ hypersurfaces, respectively.}
\label{NuCoord}
\end{figure}

Moreover, since the initial state of the detector-field system must be specified 
on the whole fiducial time-slice associated with the initial moment, if this was done by an 
observer\footnote{The ``observer" here is in the sense of relativity, who is watching nonlocally 
the quantum state of all the dynamical variables defined simultaneously on the whole time-slice associated with each moment 
in the observer's reference frame, but does not have to disturb it. It is not as restrictive as the ``observer" in QM, 
who performs measurements using specific operators or measurement devices (pointers) operating on the quantum state to 
be observed. We say that the measurement is ``local" if the operations are local in Hilbert space and 
is ``spatially local" if the operations are local in position space.} 
at rest in Minkowski space but not in past infinity, for an observer moving with constant velocity 
the initial data far enough from the detectors will appear to be specified at some times {\it after} 
the measurement event on the detector when the wavefunctional was collapsed (Figure \ref{NuCoord} (Left)).
To avoid this situation and make the quantum states comparable, we have to go beyond the linear
Lorentz transformation and inertial frames. 

A good example of the reference frame for our discussion is 
the following alternative coordinates in (1+1) dimensional Minkowski space, 
\begin{equation}
  \eta = t - A \sin t \cos x, 
  \hspace{.5cm} \xi = x - A \sin x \cos t, \label{etadef}
\end{equation}
with constant $A<1$. 
Then $ds^2=-dt^2+dx^2=\Omega(\eta, \xi)$ $\left( -d\eta^2 + d\xi^2\right)$,
where $\Omega(\eta, \xi) = (1-2A\cos t\cos x +A^2$ $\cos(t+x)\cos(t-x))^{-1}$
with $t=t(\eta, \xi)$ and $x=x(\eta, \xi)$ according to $(\ref{etadef})$. 
Here $\eta$-slices and $t$-slices will overlap at $t= \eta =n\pi$ with integer $n$,
where quantum states can be compared.
Off those moments, $\eta$-slices are the wavy ones in Figure \ref{NuCoord} (Right), where
$t$-slices would be the horizontal straight lines. 

Now the questions can be put more precisely. Suppose a pointlike detector coupled with a quantum field 
is located at $x=0$ and started at its proper time $\tau=t=0$. 
If a local measurement is done on the detector at some moment $0 < t_1 < \pi$, then which time-slice, 
a $t_1$-slice or an $\eta_1$-slice ($\eta_1\equiv\eta(t_1)$), will the wavefunctional of the combined system 
collapse on? 
If both collapses occur for different observers, will the two post-measurement states (PMS) be 
``identical"? What happens if two measurement events on two detectors are spacelike separated?

\section{Detectors in a quantum field}
\label{RSGdetector}

To answer these questions, 
let us consider one or more pointlike Raine-Sciama-Grove(RSG) detectors  \cite{RSG91} coupled to 
a massless scalar field in (1+1)D Minkowski space, described by the action
\begin{eqnarray}
  S &=& -{1\over 2}\int d^2 x \sqrt{-g} \partial_{\alpha} \Phi \partial^{\alpha} \Phi + 
    \sum_{\bf d} \left\{ {1\over 2}
    \int d\tau^{}_{\bf d} \left[ (\partial^{}_{{\bf d}} Q^{}_{\bf d})^2 -
    \omega^2_{\bf d} Q_{\bf d}^2\right] + \right. \nonumber\\ & & \left. 
    \lambda \int d\tau^{}_{\bf d} \partial^{}_{{\bf d}} Q^{}_{\bf d}
    \int d^2x \Phi(x^0,x^1) \delta^2(x^\alpha- z^\alpha_{\bf d}(\tau^{}_{\bf d})) \right\},
\label{RSGact}
\end{eqnarray} 
where $\alpha=0$ or $1$,  $g$ is the determinant of the metric tensor $g^{}_{\alpha\beta}$ of the background spacetime, 
$\Phi$ is a massless scalar field, $Q_{\bf d}$ is the internal degree of freedom acting like a harmonic oscillator (HO) 
in the detector ${\bf d}$ with ${\bf d} = A$ for the one-detector case, ${\bf d}=A,B$ for the two-detector case, and also 
$\tau^{}_{\bf d}$ and $z^\alpha_{\bf d}(\tau^{}_{\bf d})$ are the proper time and the prescribed trajectory of the detector 
${\bf d}$, respectively, and $\partial^{}_{\bf d} \equiv d/d\tau^{}_{\bf d}$. 
The momenta conjugate to $Q^{}_{\bf d}$ and $\Phi$ read
\begin{eqnarray}
  P^{}_{\bf d}(\tau_{\bf d}(x^0)) &=& {\delta S\over \delta \partial_0 Q^{}_{\bf d}} 
      = v^0_{\bf d}\left(x^0\right)
      \partial_0 Q^{}_{\bf d}\left(\tau^{}_{\bf d}(x^0)\right) + \lambda\Phi\left(x^0, z^1_{\bf d}(x^0)\right), \nonumber\\ 
  \Pi(x^0,x^1) &=& {\delta S\over \delta \partial_0 \Phi(x^0,x^1)} = -\sqrt{-g}\partial^0 \Phi(x^0,x^1),\label{Pidef}
\end{eqnarray}
where $v^0_{\bf d} = dz^0_{\bf d}/d\tau^{}_{\bf d}$. Then one can write down the Hamiltonian 
\begin{eqnarray}
  & & H(x^0) 
  = \sum_{\bf d} {1\over 2v^0_{\bf d}(x^0)}\left\{\left[ P^{}_{\bf d}(\tau_{\bf d}(x^0)) -\lambda \Phi_{z^1_{\bf d}(x^0)}(x^0)\right]^2 + 
    \omega_{\bf d}^2 Q^2_{\bf d}(\tau^{}_{\bf d}(x^0))\right\} +\nonumber\\
    & &{1\over 2}\int dx^1 \sqrt{-g}\left\{{-1\over g^{00}}\left[{\Pi_{x^1}(x^0)\over \sqrt{-g}} 
    +g^{01}\partial_1 \Phi_{x^1}(x^0)\right]^2 + 
    g^{11} \left[\partial_1\Phi_{x^1}(x^0)\right]^2\right\},
\end{eqnarray}
parameterized by the time variable $x^0$ of the observer's frame and defined on the whole time-slice $x^1 \in {\bf R}$ 
associated with that time. Solutions of the Schr\"odinger equation with the quantized Hamiltonian 
with $P^{}_{\bf d} \to \hat{P}^{}_{\bf d}  = \hbar\partial/i\partial Q^{}_{\bf d}$ 
and $\Pi^{}_{x^1} \to \hat{\Pi}^{}_{x^1}  = \hbar\delta/i\delta \Phi^{}_{x^1}$ 
are the wavefunctionals of the detectors and the field $\psi[Q^{}_{\bf d}, \Phi^{}_{x^1}; x^0]$.

Suppose that at $t_0=\eta_0=0$ (when $\tau^{}_{\bf d} \equiv 0$ for all detectors) the combined system of the 
detectors and the field is initially in a Gaussian state, which could be pure or mixed 
(e.g. the direct product of the ground states of the detectors and the Minkowski vacuum of the field.) 
Then the quantum state will always be Gaussian 
by virtue of the linearity of the system. Since the explicit form of the wavefunctional $\psi$ or the 
density matrix $\bar{\rho}[ (Q^{}_{\bf d}, \Phi^{}_{x^1}), (Q'_{\bf d}, \Phi'_{x^1}); x^0]$ is not quite simple enough to be solved directly,
we will work with the equivalent Gaussian state in the $(K,\Delta)$-representation of the density matrix \cite{UZ89}, 
which is the double Fourier transform of the conventional Wigner functional:
\begin{eqnarray}
  & &\rho[{\bf K}, {\bf \Delta}; x^0] = \int {\cal D}{\bf \Sigma} e^{{i\over\hbar} {\bf K}\cdot {\bf \Sigma}}  
    \bar{\rho}\left[ {\bf \Sigma} - {{\bf \Delta}\over 2}, {\bf \Sigma} + {{\bf \Delta}\over 2} ; x^0 \right] 
    \nonumber\\  &=& \exp -{1\over 2\hbar^2} \left[ K^\mu {\cal Q}_{\mu\nu}(x^0) K^\nu 
    -2 \Delta^\mu {\cal R}_{\mu\nu}(x^0) K^\nu + \Delta^\mu {\cal P}_{\mu\nu}(x^0) \Delta^\nu \right],
\label{Qstate}
\end{eqnarray}
where DeWitt notation has been used, $\mu, \nu \in \{{\bf d}\}\cup\{ x^1\}$ run over 
all the detector and field degrees of freedom defined on the whole time-slice, and the time-dependent 
factors ${\cal Q}_{\mu\nu}(x^0)$, ${\cal P}_{\mu\nu}(x^0)$, and ${\cal R}_{\mu\nu}(x^0)$ are actually the symmetric 
two-point correlators $\langle A,B\rangle \equiv \langle AB+BA \rangle/2$ evaluated
on the $x^0$-slice, for they are obtained from 
\begin{eqnarray}
  \langle \hat{\Phi}^{}_\mu , \hat{\Phi}^{}_\nu \rangle 
  &=& \left. {\hbar\delta\over i\delta K^\mu}{\hbar\delta\over i\delta K^\nu} 
    \rho[{\bf K}, {\bf \Delta}; x^0] \right|_{{\bf \Delta}={\bf K}=0} = {\cal Q}_{\mu\nu},\nonumber\\
  \langle \hat{\Pi}^{}_\mu , \hat{\Pi}^{}_\nu \rangle &=&
    \left. {i\hbar\delta\over \delta \Delta^\mu}{i\hbar\delta\over \delta \Delta^\nu} 
    \rho[{\bf K}, {\bf \Delta}] \right|_{{\bf \Delta} ={\bf K}=0} = {\cal P}_{\mu\nu},\nonumber\\
  \langle \hat{\Pi}^{}_\mu , \hat{\Phi}^{}_\nu \rangle &=& 
    \left. {i\hbar\delta\over \delta \Delta^\mu}{\hbar\delta\over i\delta K^\nu} 
    \rho[{\bf K}, {\bf \Delta}] \right|_{{\bf \Delta} = {\bf K}=0} = {\cal R}_{\mu\nu},
\end{eqnarray}
where we denote $\hat{Q}_{\bf d}$ and $\hat{P}_{\bf d}$ by $\hat{\Phi}^{}_{\bf d}$ and $\hat{\Pi}^{}_{\bf d}$,
respectively. Thus, looking at the evolution of the Gaussian state $(\ref{Qstate})$ is equivalent to 
looking at the dynamics of those symmetric two-point correlators, which would be obtained
more easily in the Heisenberg picture. 

\begin{figure}
\includegraphics[width=3.95cm]{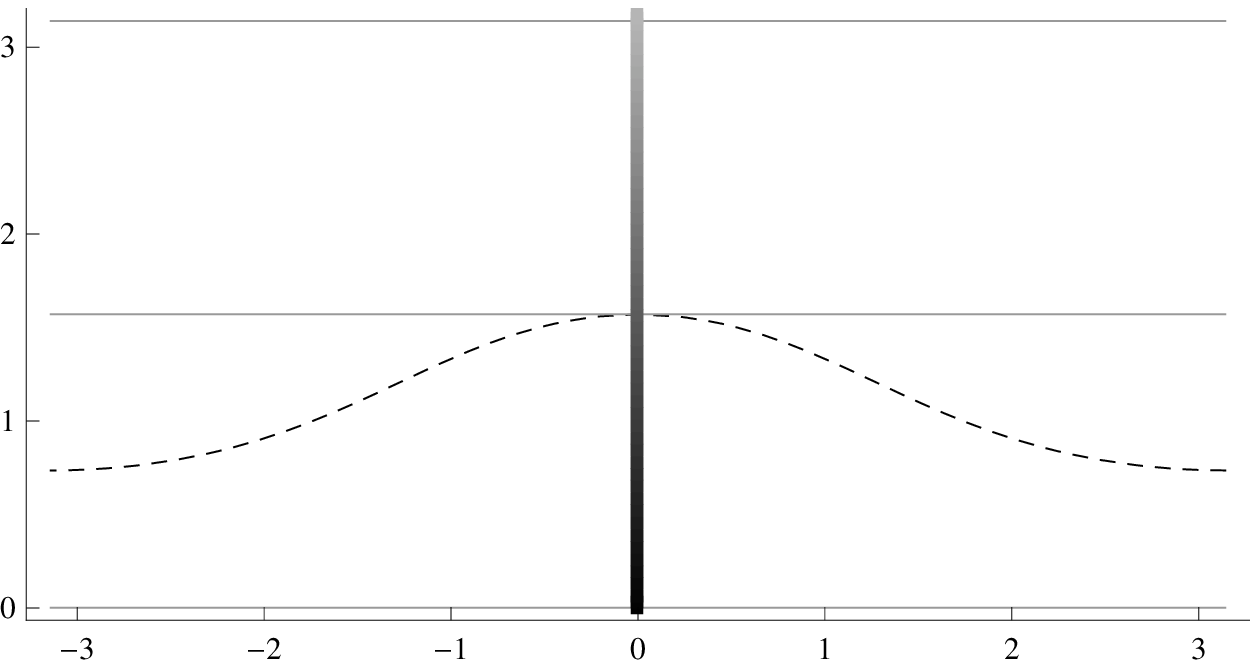}
\includegraphics[width=3.95cm]{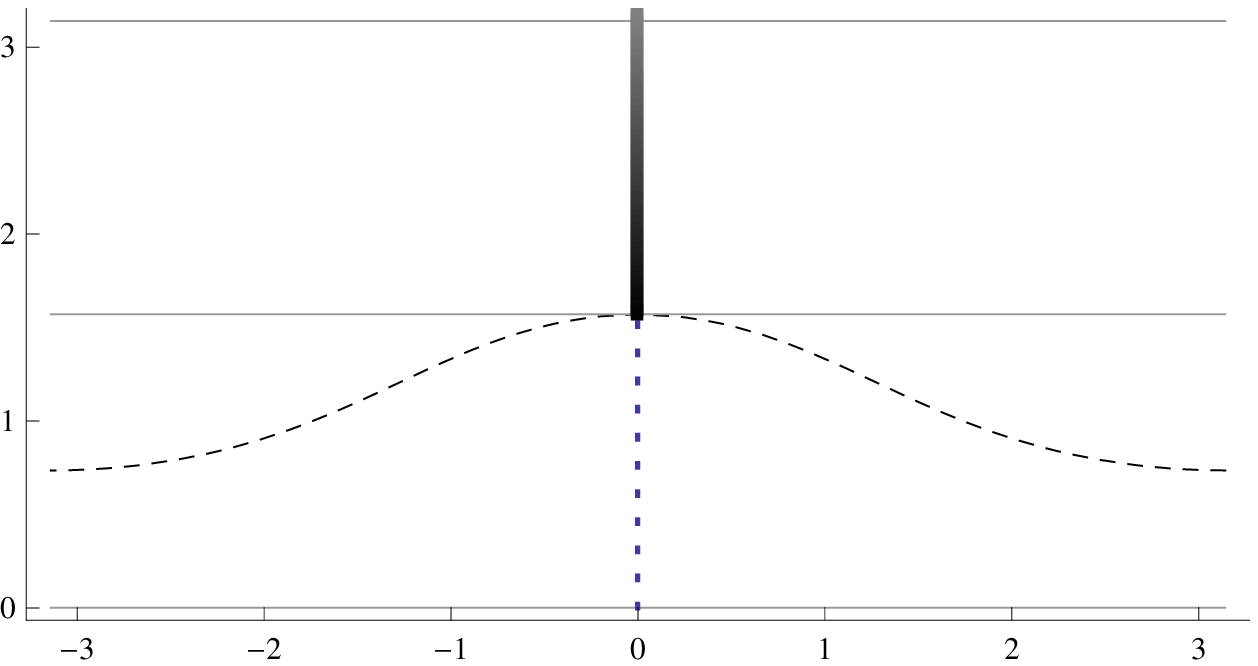}
\includegraphics[width=3.95cm]{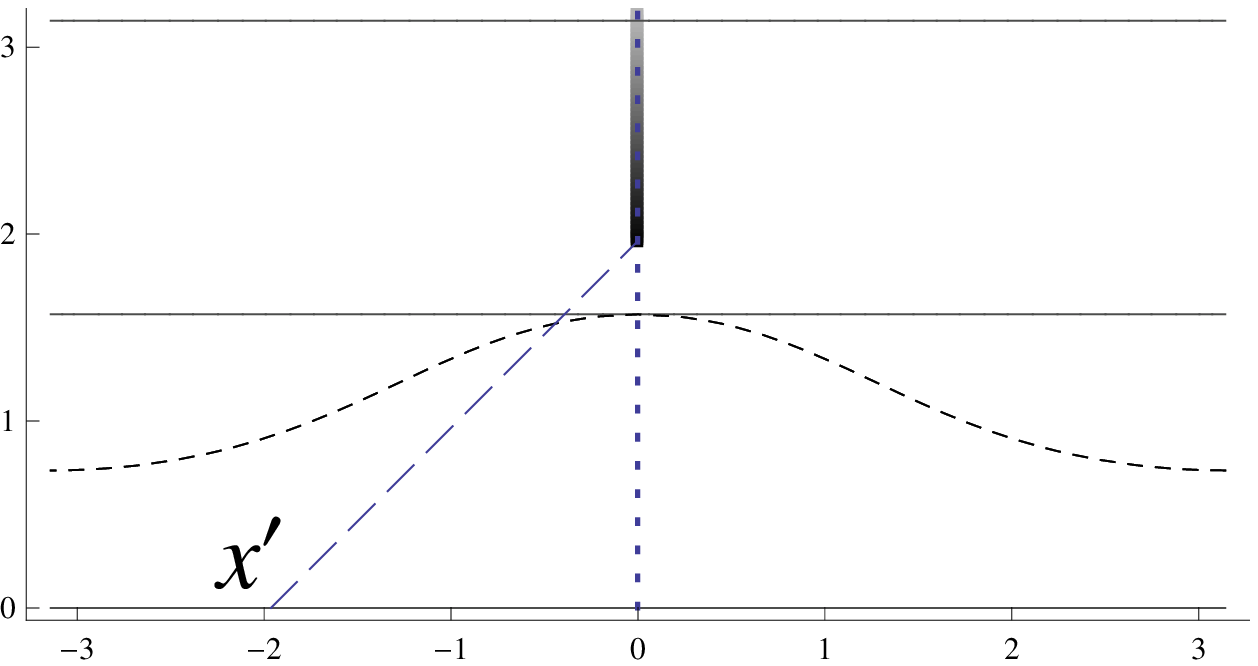}\\
\includegraphics[width=3.95cm]{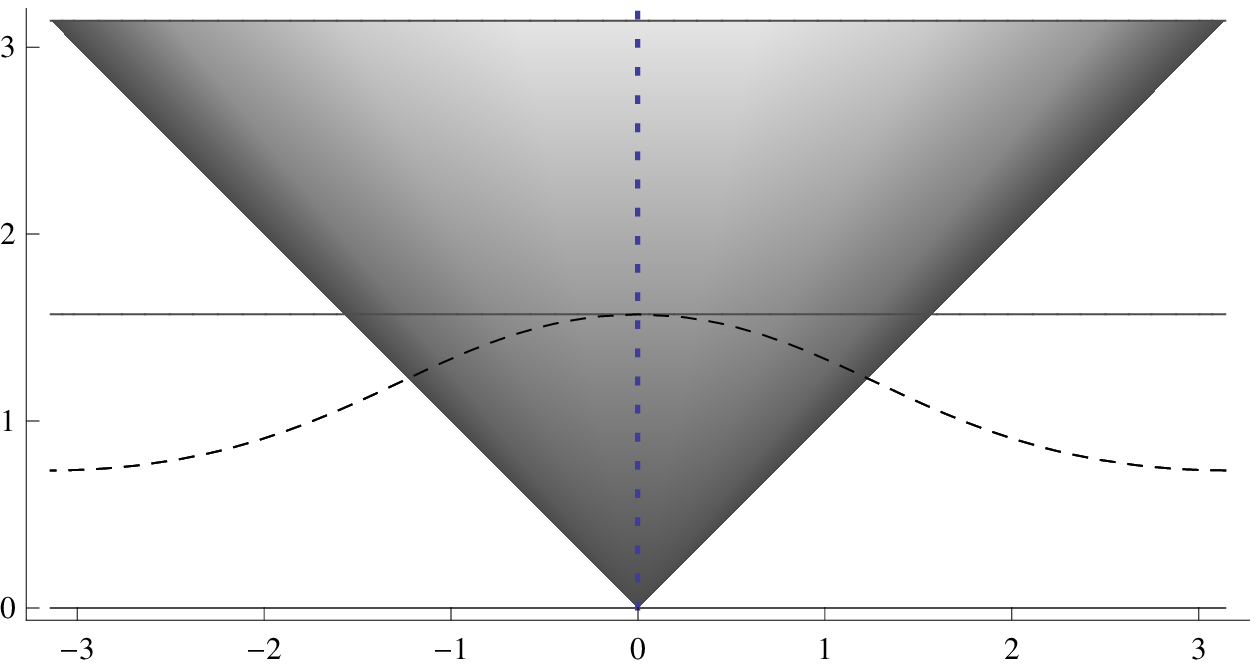}
\includegraphics[width=3.95cm]{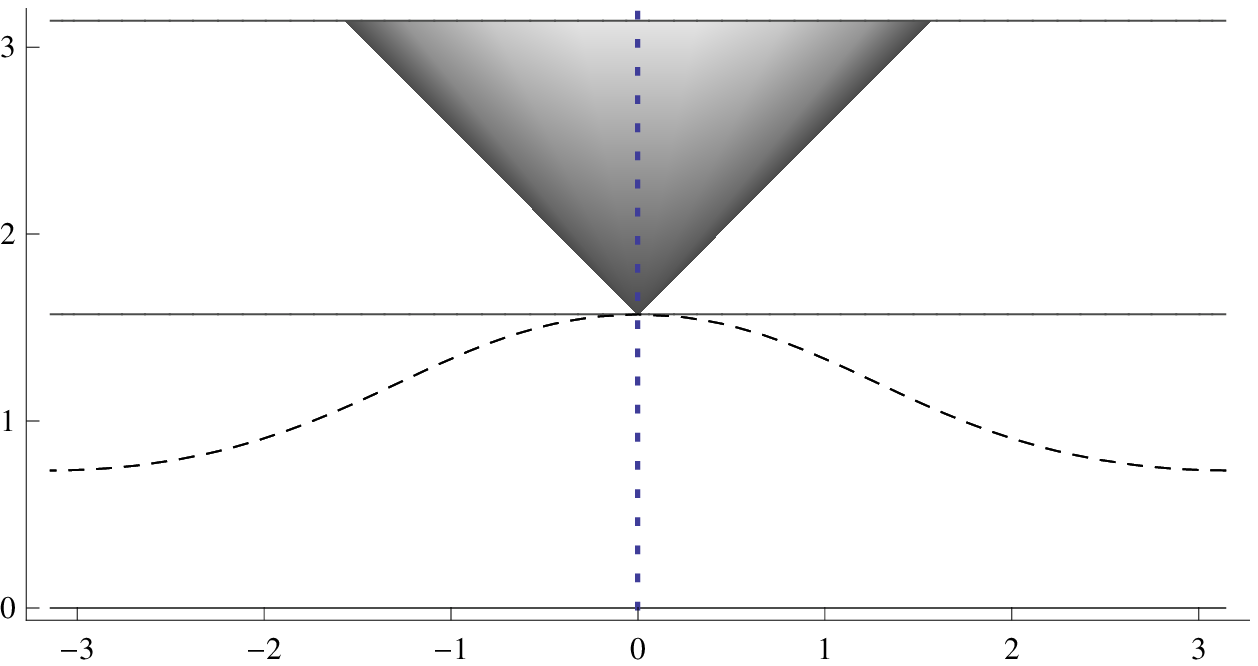}
\includegraphics[width=3.95cm]{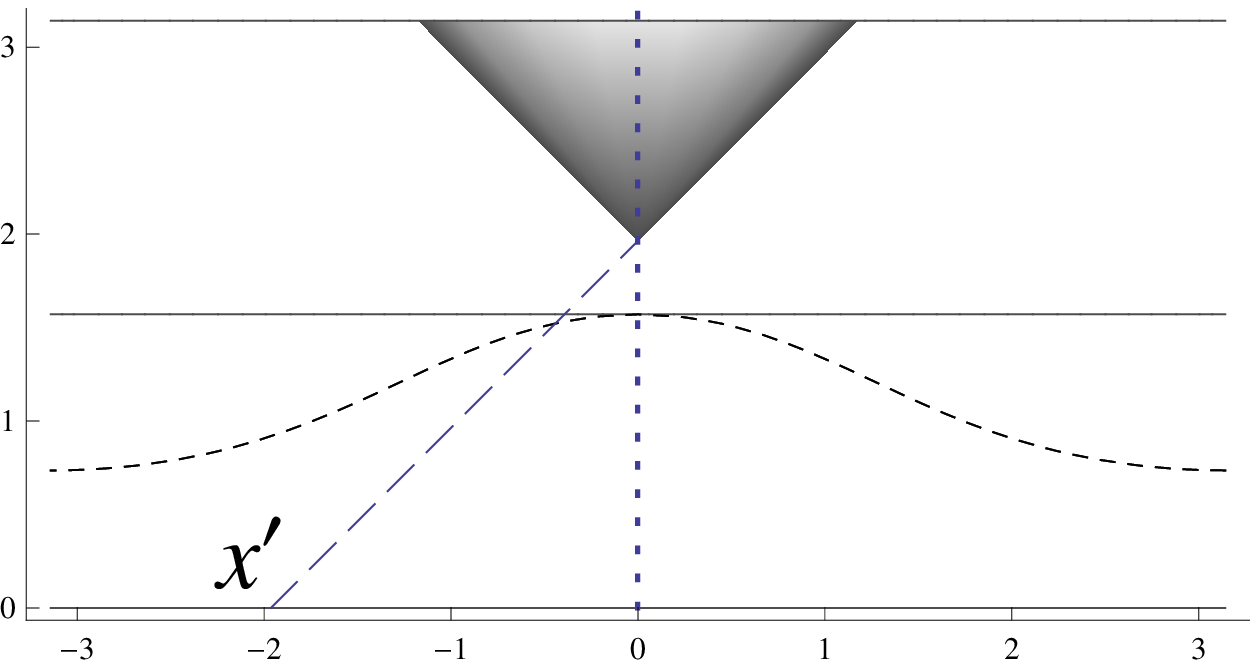}
\caption{The diagrams in the upper row represent the mode functions of the oscillators inside the detector, 
(from left to right) $\phi_A^{A[20]}$, $\phi_A^{A[21]}$, and $\phi_A^{x'[20]}$,
which generate the retarded fields $\phi_x^{A[20]}$, $\phi_x^{A[21]}$, and $\phi_{x}^{x'(1)[20]}$ through
$(\ref{fieldxx1})$ and $(\ref{phiinho})$ represented
in the diagrams from left to right in the lower row, respectively. 
Here the gray horizontal lines denote the $t$-slices at $t=\eta=0$, $t=t_1=\pi/2$, and $t=t_2=\eta_2=\pi$, the dashed 
wavy lines are $\eta$-slices with $\eta = \eta_1=(\pi-1)/2$ (here we choose $A=1/2$ in $(\ref{etadef})$), 
and the dotted vertical lines are the worldlines of detector $A$ at $x=\xi=0$.   
All the above mode functions are independent of the data on the $t_1$- or $\eta_1$-slices except those right at the 
position of the pointlike detector $(t,x)=(\pi/2,0)$ where the local measurement was done.
The long-dashed lines in the two plots for $\phi_A^{x'[20]}$ (upper right) and $\phi_x^{x'(1)[20]}$ (lower right) 
denote the vacuum fluctuations started from $x=x'$ at $t=0$ (cf.   $(\ref{zerophi})$).} 
\label{phidiag}
\end{figure}

The Heisenberg equations of motion for the operators $\hat{Q}^{}_{\bf d}$ and $\hat{\Phi}(x^1)$ read
\begin{eqnarray}
  & &\partial^{}_{\bf d} \hat{Q}^{}_{\bf d}(\tau^{}_{\bf d}) + \omega_{\bf d}^2 \hat{Q}^{}_{\bf d}(\tau^{}_{\bf d}) 
    = -\lambda \partial^{}_{\bf d} \hat{\Phi}(z^\alpha_{\bf d}(\tau^{}_{\bf d})), \label{HOeom} \\
  & &\sqrt{-g}\Box\hat{\Phi}(x^\alpha) = \lambda \sum_{\bf d} \int d\tau^{}_{\bf d} \partial^{}_{\bf d} \hat{Q}^{}_{\bf d} 
    \delta^2\left(x^\alpha-z^\alpha_{\bf d}(\tau^{}_{\bf d})\right) \label{fieldeq}
\end{eqnarray}
where $\Box\equiv \sqrt{-g}^{-1}\partial_{\alpha} \sqrt{-g}g^{\alpha\beta}\partial_{\beta} $.
By virtue of the linearity of the system, operators at each moment 
are linear combinations of the operators defined at the initial moment \cite{LH06}:
\begin{eqnarray}
  \hat{Q}^{}_{\bf d}(\tau^{}_{\bf d}) &=& \sum_{{\bf d}'} 
    \left[\phi^{{\bf d}'}_{\bf d}(\tau^{}_{\bf d})\hat{Q}^{[0]}_{{\bf d}'} + 
    f^{{\bf d}'}_{\bf d}(\tau^{}_{\bf d})\hat{P}^{[0]}_{{\bf d}'} \right] \nonumber\\ & &
    +\int dy \left[ \phi^{y}_{\bf d}(\tau^{}_{\bf d})\hat{\Phi}^{[0]}_{y} + 
    f^{y}_{\bf d}(\tau^{}_{\bf d})\hat{\Pi}^{[0]}_{y} \right], \label{Qexp}\\
  \hat{\Phi}^{}_{x^1}(x^0) &=&
    \sum_{{\bf d}'} \left[\phi^{{\bf d}'}_{x^1}(x^0)\hat{Q}^{[0]}_{{\bf d}'} +
    f^{{\bf d}'}_{x^1}(x^0)\hat{P}^{[0]}_{{\bf d}'}\right] \nonumber\\ & & 
    +\int dy \left[ 
    \phi^{y}_{x^1}(x^0)\hat{\Phi}^{[0]}_{y} + f^{y}_{x^1}(x^0)\hat{\Pi}^{[0]}_{y} \right], \label{Phiexp}
\end{eqnarray}
from which $\hat{P}^{}_{\bf d}(\tau_{\bf d})$ and $\hat{\Pi}^{}_{x^1}(x^0)$ can be derived according to $(\ref{Pidef})$.
Here $\hat{\cal O}_{\mu}^{[n]} \equiv \hat{\cal O}_{\mu}(t_n)$ and
all the ``mode functions" $\phi^{\mu}_\nu(x^0)$ and $f^{\mu}_\nu(x^0)$ are real functions of time,
which can be related to those in $k$-space in   \cite{LH06} (with different initial conditions, though.)
Inserting the above expansions into $(\ref{fieldeq})$, one has
\begin{equation}
  \sqrt{-g}\Box\phi^\mu_{x^1}(x^0) = \lambda \sum_{\bf d} \int d\tau^{}_{\bf d} \partial^{}_{\bf d} \phi^\mu_{\bf d}
  \delta^2\left(x^\alpha-z^\alpha_{\bf d}(\tau^{}_{\bf d})\right),
\label{fieldxx1}
\end{equation}
which gives $\phi^\mu_{x^1}(x^0)=\phi^{\mu(0)}_{x^1}(x^0)+\phi^{\mu(1)}_{x^1}(x^0)$ where,
for proper initial conditions, the homogeneous solutions are 
\begin{eqnarray} 
  \phi^{{\bf d} (0)}_{x^1}(x^0) &=& 0, \label{zeroq} \\
  \phi^{y^1(0)}_{x^1}(T) &=& \int {dk\over 2\pi} e^{ik(x^1-y^1)}\cos\omega_k T \nonumber\\ &=& 
  {1\over 2}\left[ \delta(x^1-y^1 + T)+\delta(x^1-y^1-T)\right], \label{zerophi}
\end{eqnarray}
with $\omega_k = |k|$, while the inhomogeneous solutions read
\begin{equation}
  \phi^{\mu(1)}_{x^1}(x^0) =  \sum_{{\bf d}} \lambda \int_0^\infty 
  d\tau^{}_{\bf d} G_{\rm ret}(x^\alpha; 
  z^\alpha_{\bf d}(\tau^{}_{\bf d})) \partial^{}_{\bf d} \phi^{\mu}_{{\bf d}}(\tau^{}_{\bf d})
\label{phiinho}
\end{equation}
with the retarded Green's function for the massless scalar field in (1+1)D, 
$ G_{\rm ret}\left(t,x; t',x'\right) =
\theta\left[ t+x -(t'+x') \right]$ $\theta\left[ t-x -(t'-x') \right]/2$
in the $t$-$x$ frame. Now $\phi^{y^1(0)}_{x^1}(x^0)$ can be interpreted as vacuum fluctuations of the field
propagating from $(0, y^1)$ to $(x^0,x^1)$, 
while $\phi^{\mu(1)}_{x^1}(x^0)$ behave like retarded relativistic fields sourced by the pointlike detectors
$\phi^{\mu}_{{\bf d}}(\tau^{}_{\bf d})$.
Inserting the solutions of $\phi^\mu_{x^1}$ into $(\ref{HOeom})$, one obtains
\begin{eqnarray}
  & & \left( \partial_{\bf d}^2 + 2\gamma\partial_{\bf d} +\omega_{\bf d}^2\right) \phi^{\mu}_{\bf d}(\tau^{}_{\bf d})
   = -\lambda \partial^{}_{\bf d} \phi_{z^1_{\bf d}(\tau^{}_{\bf d})}^{\mu(0)}\left(z^0_{\bf d}(\tau^{}_{\bf d})\right) -
   \nonumber\\ & & \hspace{1.5cm} 
   4\gamma \partial^{}_{{\bf d}} \sum_{{\bf d}'\not={\bf d}}\int_0^\infty 
    d\tau^{}_{{\bf d}'} G_{\rm ret}\left(z^\alpha_{{\bf d}}(\tau^{}_{{\bf d}}); 
  z^\alpha_{{\bf d}'}(\tau^{}_{{\bf d}'})\right)   
  \partial^{}_{{\bf d}'} \phi^{\mu}_{{\bf d}'}(\tau^{}_{{\bf d}'})
   \label{eomHO1}
\end{eqnarray}
with $\gamma\equiv\lambda^2/4$. For the cases with a single detector, the right hand side of $(\ref{eomHO1})$ for
$\phi^{{\bf d}'}_{\bf d}$ is simply zero.
From the equations of motion $(\ref{eomHO1})$ one learns that $\phi^{{\bf d}'}_{\bf d}$ 
behave like damped HOs, while $\phi^{y^1}_{\bf d}$ behave like damped HOs driven by vacuum 
fluctuations of the field $\phi^{y^1(0)}_{z^1_{{\bf d}}(\tau^{}_{{\bf d}})}$ at the position of the detector. 
Both are living in the pointlike detectors and not extended in space.
Both would also be driven by retarded mutual influences from the other detectors, 
if the last term of $(\ref{eomHO1})$ is non-vanishing. In Figure \ref{phidiag} these mode functions
are represented in diagrams. 

\section{One-detector case} 

Suppose an RSG detector (detector $A$) is at rest in Minkowski space with the worldline $z_A^\alpha = (t,0)$ in
the $t$-$x$ frame and $(\eta,0)$ in the $\eta$-$\xi$ frame, and the detector-field system is initially in a Gaussian state
at $t=\eta=0$. Suppose a Gaussian measurement is done on detector $A$ at 
$t=t_1 \in (0, \pi)$ when the quantum state on the $t_1$-slice collapses to 
\begin{equation}
  \tilde{\rho} = \rho^{}_A \otimes \rho^{}_{\bar{A}}, \label{rhoPMS}
\end{equation}
for the observer in the $t$-$x$ frame, such that 
\begin{equation}
  \rho^{}_A = \exp -{1\over 4\hbar} \left[ g_A \left(\Delta^A\right)^2 + 
  {1\over g_A}\left(K^A\right)^2\right] \label{rhoA}
\end{equation}
is a Gaussian state of detector $A$ with $\langle(\hat{Q}^{[1]}_A)^2\rangle_1 = 
\hbar g_A /2$, $\langle(\hat{P}^{[1]}_A)^2\rangle_1 = \hbar/2g_A$, and
$\langle\hat{Q}_{A}^{[1]},\hat{P}_A^{[1]}\rangle_1=0$ with some constant $g^{}_A$.
Here $\langle \,\, \rangle_n$ denotes the expectation values 
taken from the quantum state right after $t=t_n$. From $(\ref{rhoA})$ one obtains 
\begin{eqnarray}
  \rho^{}_{\bar{A}} &=& N\int {dK^A d\Delta^A\over 2\pi\hbar} \rho_A^* \cdot \rho[{\bf K},{\bf \Delta}; t_1]
    \nonumber\\ &=& \exp -{1\over 2\hbar^2}
    \left[ K^{\bar{\mu}} \tilde{\cal Q}_{\bar{\mu}\bar{\nu}}  K^{\bar{\nu}} - 
    2\Delta^{\bar{\mu}} \tilde{\cal R}_{{\bar{\mu}}{\bar{\nu}}} K^{\bar{\nu}} + 
    \Delta^{\bar{\mu}} \tilde{\cal P}_{{\bar{\mu}}{\bar{\nu}}} \Delta^{\bar{\nu}} \right],
\end{eqnarray}
for the rest of the system. Here $N$ is the normalization factor, $\bar{\mu}, \bar{\nu} \in \{x\}\cup\{{\bf d}\}-\{A\}$,
$\tilde{\cal Q}_{\bar{\mu}\bar{\nu}} =\langle \hat{\Phi}_{\bar{\mu}}^{[1]} , \hat{\Phi}_{\bar{\nu}}^{[1]} \rangle_{1}$, 
$\tilde{\cal R}_{\bar{\mu}\bar{\nu}} = \langle \hat{\Phi}_{\bar{\mu}}^{[1]} , \hat{\Pi}_{\bar{\nu}}^{[1]} \rangle_{1}$, and
$\tilde{\cal P}_{\bar{\mu}\bar{\nu}} = \langle \hat{\Pi}_{\bar{\mu}}^{[1]},$ $\hat{\Pi}_{\bar{\nu}}^{[1]} \rangle_{1}$,
where
\begin{equation}
  \langle \hat{\Theta}_{\bar{\mu}\, ,}^{[1]}  \hat{\Theta}_{\bar{\nu}}^{[1]} \rangle_{1} =
  \langle \hat{\Theta}_{\bar{\mu}\; ,}^{[10]} \hat{\Theta}_{\bar{\nu}}^{[10]} \rangle_{0} +                            
  {I^{[1,0]}_A (\hat{\Theta}_{\bar{\mu}\; ,}^{[10]} \hat{\Theta}_{\bar{\nu}}^{[10]} )\over J_A^{[1,0]}}.
  \label{cort1t0}
\end{equation}
with $\Theta = \Phi$ or $\Pi$, ${\cal O}^{[ns]}_{\mu}\equiv {\cal O}^{}_\mu(t_n-t_s)$, and
\begin{eqnarray}
  & & I^{[n,s]}_{\bf d} \left(\hat{\cal O},\hat{\cal O}'\right) \equiv \nonumber\\ & &   
  -\langle \hat{\cal O}_, \hat{Q}_{{\bf d}}^{[ns]} \rangle_{s}  
    \langle\hat{\cal O'}_, \hat{Q}_{{\bf d}}^{[ns]} \rangle_{s}\ell^{[1,0]}_{\bf d} 
  -\langle\hat{\cal O}_, \hat{P}_{{\bf d}}^{[ns]} \rangle_{s}  
  \langle \hat{\cal O'}_, \hat{P}_{{\bf d}}^{[ns]} \rangle_{s}\bar{\ell}^{[1,0]}_{\bf d} 
  \nonumber\\ && + \langle \hat{Q}_{{\bf d}\; ,}^{[ns]} \hat{P}_{{\bf d}}^{[ns]} \rangle_{s} 
  \left[ \langle \hat{\cal O}_, \hat{Q}_{{\bf d}}^{[ns]} \rangle_{s}  
    \langle \hat{\cal O'}_, \hat{P}_{{\bf d}}^{[ns]} \rangle_{s}
    +(\hat{\cal O}\leftrightarrow \hat{\cal O}') \right],
\end{eqnarray}
with $\ell^{[1,0]}_{\bf d} \equiv \langle (\hat{Q}_{\bf d}^{[1]})^2\rangle_1 +\langle 
(\hat{P}^{[10]}_{\bf d})^2\rangle_0$, $\bar{\ell}^{[1,0]}_{\bf d} \equiv \langle(\hat{P}_{\bf d}^{[1]})^2
\rangle_1+$ $\langle(\hat{Q}^{[10]}_{\bf d})^2\rangle_0$, and $J_{\bf d}^{[n,s]} \equiv 
\ell^{[n,s]}_{\bf d} \bar{\ell}^{[n,s]}_{\bf d} -\langle\hat{Q}^{[ns]}_{\bf d\; ,}\hat{P}^{[ns]}_{\bf d}\rangle_s^2 $. 
For the observer in the $\eta$-$\xi$ frame, according to the postulate of the projective measurement in that frame, we assume 
that a similar projection occurs but the wavefunctional collapses on the $\eta_1$-slice instead. 
So the PMS and the factors therein have the same form as the above ones in the $t$-$x$ frame except that the correlators are 
evaluated in the $\eta$-$\xi$ frame. Then, starting at $t_1$ and $\eta_1$, both the PMS in the $t$-$x$ frame
and the PMS in the $\eta$-$\xi$ frame evolve to $t_2=\pi=\eta_2$, when $t$ and $\eta$-slices overlap and two observers 
can make a comparison on these two quantum states. 

In the conventional $(t, x)$ coordinates of Minkowski space, the two-point correlators at $t_2$ determining 
the wavefunctional can be expressed as combinations of the mode functions evolving 
from $t_1$ to $t_2$, together with the initial data on the $t_1$-slice in the form of the correlators of the 
field at space points on the slice, e.g. from $(\ref{Qexp})$ and $(\ref{Phiexp})$,
\begin{eqnarray}
  \tilde{\cal Q}_{xy}(t_2) &=& \langle \hat{\Phi}_x^{[2]}, \hat{\Phi}_y^{[2]} \rangle_{2} =
  {\rm Tr} \hat{\Phi}_x^{[21]} \hat{\Phi}_y^{[21]} \tilde{\rho} \nonumber\\ 
  &\sim&  \int dx' dy' \phi_x^{x'[21]}\phi_y^{y'[21]} \langle \hat{\Phi}_{x'}^{[1]}, 
  \hat{\Phi}_{y'}^{[1]} \rangle_{1} + \cdots , \label{Qxyt1}
\end{eqnarray}
where $x'$ and $y'$ are points on the $t_1$-slice. Apparently $\langle \hat{\Phi}_{x'}^{[1]}, \hat{\Phi}_{y'}^{[1]} 
\rangle_{1}$ depends on the data on the $t_1$-slice, and as mentioned below $(\ref{fieldxx1})$,  $\phi_x^{x'[21]}$ 
is a superposition of vacuum fluctuations $\phi_x^{x'(0)[21]}$  propagating from the point $(t_1, x')$ on the $t_1$-slice
to $(t_2, x)$ and the retarded field $\phi_x^{x'(1)[21]}$ sourced by the pointlike detector driven by vacuum 
fluctuations from the $t_1$-slice.
On the other hand, in the alternative coordinates $(\eta,\xi)$, the form of 
the correlator is similar to $(\ref{Qxyt1})$, except that the dependence is on the $\eta_1$-slice.
So here the two-point correlators, or equivalently,
the wavefunctionals at $t_2=\eta_2$, appear to depend on the time-slicing scheme.  

Nevertheless, by comparing the expansions $(\ref{Qexp})$ and $(\ref{Phiexp})$ of two equivalent evolutions
without considering any measurement:
one from $t_0$ to $t_1$ then from $t_1$ to $t_2$, the other from $t_0$ all the way to $t_2$,
one can see that the mode functions have the following identities,
\begin{eqnarray}
  \phi^{\mu[20]}_\nu &=& 
    \phi_\nu^{\sigma [21]}\phi_{\sigma}^{\mu[10]} + 
    f_\nu^{\sigma [21]} \pi^{\mu[10]}_{\sigma},\label{id1}\\ 
  f^{\mu[20]}_\nu  &=& 
    \phi_\nu^{\sigma [21]}f_{\sigma}^{\mu[10]} +
    f_\nu^{\sigma [21]} p^{\mu[10]}_{\sigma} \label{id2} 
\end{eqnarray}
where DeWitt notation is understood, 
$\pi^{\mu}_{\bf d}(\tau^{}_{\bf d}(t)) \equiv \partial^{}_{\bf d}\phi^{\mu}_{{\bf d}}
(\tau^{}_{\bf d}(t))+ \lambda\phi^{\mu}_{x^1_{\bf d}(t)}(t)$, $\pi^{\mu}_x(t) \equiv \partial_t \phi^{\mu}_x(t)$,
$p^{\mu}_{\bf d}(\tau^{}_{\bf d}(t)) \equiv \partial^{}_{\bf d}f^{\mu}_{{\bf d}}
(\tau^{}_{\bf d}(t))+ \lambda f^{\mu}_{x^1_{\bf d}(t)}(t)$ and $p^{\mu}_x(t) \equiv \partial_t f^{\mu}_x(t)$ 
with $\mu, \nu, \sigma \in \{{\bf d}\}\cup \{x^1\}$ according to $(\ref{Pidef})$. 
Similar identities for $\pi^\mu_\nu$ and $p^\mu_\nu$ can be derived straightforwardly from $(\ref{id1})$ and $(\ref{id2})$. 
Such identities can be interpreted as the Huygens principle of the mode functions, 
and can be verified by inserting particular solutions of the mode functions into the identities.

The dependence of the $t_1$-slice in $(\ref{Qxyt1})$ turns out to be removable by expressing the $\langle \ldots \rangle_1$ 
correlators as $(\ref{cort1t0})$ with $(\ref{Qexp})$ and $(\ref{Phiexp})$ inserted, then using $(\ref{id1})$, $(\ref{id2})$ 
and similar identities for $\pi^\mu_\nu$ and $p^\mu_\nu$ to replace the $\int dx'$ and $\int dy'$ terms to reach 
\begin{eqnarray}
  \tilde{\cal Q}_{xy}(t_2) &=&
  {\hbar\over 2}g^{}_A \phi_x^{A[21]} \phi_y^{A[21]}+{\hbar\over 2 g^{}_A} f_x^{A[21]} f_y^{A[21]} \nonumber\\ & &
  + \langle\hat{\Upsilon}_x^{[0]}, \hat{\Upsilon}_y^{[0]}\rangle_0 
  + {1\over J_A^{[1,0]}} I_A^{[1,0]} \left(\hat{\Upsilon}_x^{[0]}, \hat{\Upsilon}_y^{[0]} \right)
\end{eqnarray}
where 
\begin{eqnarray}
  \hat{\Upsilon}_x^{[0]} &\equiv& \hat{\Phi}_{\mu}^{[0]} \left(\phi_x^{\mu[20]}-\phi_x^{A[21]}\phi_A^{\mu[10]} - 
    f_x^{A[21]}\pi_A^{\mu[10]}\right) + \nonumber\\& & 
  \hat{\Pi}_{\mu}^{[0]}\left(f_x^{\mu[20]}-\phi_x^{A[21]}f_A^{\mu[10]} - f_x^{A[21]}p_A^{\mu[10]}\right). 
\end{eqnarray} 
Then all the two-point correlators, and therefore the 
$(K, \Delta)$-representation of the quantum state on the $t_2$-slice, end up with functionals of $F_x^{\mu[20]}$, 
$F_x^{A[21]}$, $F_A^{\mu[20]}$, $F_A^{\mu[10]}$, $F_A^{A[21]}$ with $F = \phi, f, \pi, p$,  
and the initial data in the two-point correlators evaluated at $t_0$.

The quantum state at $\eta_2$ in the $\eta$-$\xi$ frame has exactly the same functional form. 
Now $F_A^{\mu[20]}$, $F_A^{\mu[10]}$, and $F_A^{A[21]}$ are HOs in the pointlike detector, $F_x^{A[21]}$ and 
$F_x^{A[20]}$ are retarded fields sourced from the detector, while $F_x^{x'[20]}$ are superpositions of vacuum 
fluctuations and retarded fields, as indicated from $(\ref{fieldxx1})$ to $(\ref{eomHO1})$. 
All of them are explicitly covariant and independent of the data on the $t_1$- or $\eta_1$-slice outside the detector
(see Figure \ref{phidiag}).  Thus the quantum states collapsed in two different frames are identical up to a 
coordinate transformation when compared on the same time-slice at $t_2=\eta_2=\pi$.

The case with two successive projective measurements on a detector at $t=t_1$ and $t_2$, $0<t_1 < t_2 < t_3=\pi$ 
has also been calculated. The two-point correlators at $t_3$ can similarly be written in functionals of the 
covariant functions $F_x^{x'[30]}$, $F_A^{x'[m0]}$, $F_x^{A[3n]}$, 
$F_A^{A[mn]}$ with $3\ge m>n \ge 0$, and the initial data at $t_0$. They are independent of the data on 
$t_1$-, $\eta_1$-, $t_2$-, or $\eta_2$-slice outside the detector, so 
the quantum state at $t_3$ is still independent of the time-slices on which the wavefunctional collapsed.
It is straightforward to generalize this result to the cases with many successive projective measurements 
on the detector.

\section{Two-detector case}

Now consider the case with two spatially separated detectors. For simplicity, $Q_A$ is put at $x=-\pi$ and 
$Q_B$ at $x=0$, so both are at rest in the $t$-$x$ and $\eta$-$\xi$ frames. Suppose a local measurement like $(\ref{rhoPMS})$
is done on $Q_A$ at $t_1$ and $\eta_1=t_1 + A\sin t_1$, and a similar local measurement on $Q_B$ at $t_2>t_1$ but 
$\eta_2 = t_2 -A \sin t_2<\eta_1$: the two measurement events are spacelike separated, so the time order of these two events can 
be altered. In this case, the two-point functions at $t_3=\eta_3=\pi$ can still be written in functionals independent of the 
data on the $t_1$-, $\eta_1$-, $t_2$-, or $\eta_2$-slice outside the detectors. For example, in the $t$-$x$ frame,
\begin{eqnarray}
  & &\tilde{\cal Q}_{xy}(t_3) =\langle \hat{\Phi}_x^{[3]}, \hat{\Phi}_y^{[3]} \rangle_{3}
  \nonumber\\ &=& {\hbar\over 2}g^{}_B \phi_x^{B[32]} \phi_y^{B[32]}
  +{\hbar\over 2 g^{}_B} f_x^{B[32]} f_y^{B[32]} 
  +{\hbar\over 2}g^{}_A \phi_x^{A[31]} \phi_y^{A[31]}+{\hbar\over 2 g^{}_A} f_x^{A[31]} f_y^{A[31]} \nonumber\\ & &
  + \langle\hat{\Upsilon}_x^{[0]}, \hat{\Upsilon}_y^{[0]}\rangle_0 
  + {1\over J_A^{[1,0]}} I_A^{[1,0]} \left(\hat{\Upsilon}_x^{[0]}, \hat{\Upsilon}_y^{[0]} \right)
  + {1\over J_B^{[2,1]}} I_B^{[2,1]} \left(\hat{\Upsilon}_x^{[1]}, \hat{\Upsilon}_y^{[1]} \right),
\end{eqnarray}
with $\hat{\Upsilon}_x^{[n]} \equiv \hat{\Phi}_{\mu}^{[n]}\tilde{\phi}_x^{\mu[3n]} +\hat{\Pi}_{\mu}^{[n]}\tilde{f}_x^{\mu[3n]}$ and
\begin{eqnarray}
  \tilde{\phi}_x^{\mu[30]} \equiv \phi_x^{\mu[30]}-\phi_x^{A[31]}\phi_A^{\mu[10]} - 
   f_x^{A[31]}\pi_A^{\mu[10]}-\phi_x^{B[32]}\phi_B^{\mu[20]}- f_x^{B[32]}\pi_B^{\mu[20]}, \\
  \tilde{f}_x^{\mu[30]} \equiv f_x^{\mu[30]}-\phi_x^{A[31]}f_A^{\mu[10]} - f_x^{A[31]}p_A^{\mu[10]}
    -\phi_x^{B[32]}f_B^{\mu[20]} - f_x^{B[32]}p_B^{\mu[20]},\\ 
  \tilde{\phi}_x^{\mu[31]}  \equiv \phi_x^{\mu[31]}-\phi_x^{B[32]}\phi_B^{\mu[21]}- f_x^{B[32]}\pi_B^{\mu[21]}, \\  
  \tilde{f}_x^{\mu[31]} \equiv f_x^{\mu[31]}-\phi_x^{B[32]}f_B^{\mu[21]} - f_x^{B[32]}p_B^{\mu[21]}. 
\end{eqnarray}
On expressing all the correlators $\langle\ldots\rangle_1$ in $J_B^{[2,1]}$ and $I_B^{[2,1]}$ in terms of 
$\langle\ldots\rangle_0$ with the help of $(\ref{cort1t0})$, both $J_B^{[2,1]}$ and $I_B^{[2,1]}(\hat{\Upsilon}_x^{[1]}, 
\hat{\Upsilon}_y^{[1]})$ will become functionals of 
$\tilde{F}_x^{\mu[30]}$, $F_B^{\mu[20]}$, and the correlators evaluated on the initial time-slice. 
So the final expression for the wavefunctional at $t_3$ depends only 
on the covariant mode functions corresponding to the HOs in the detectors, the retarded fields sourced by 
detectors A and B, vacuum fluctuations, as well as the initial data on the $t_0$-slice. 
Further, using computer algebra it is straightforward to verify that whenever the retarded mutual influence $F_B^{A[21]}$ is zero,  
$I_A^{[1,2]} (\hat{\Upsilon}_{\xi}^{[2]}, \hat{\Upsilon}_{\xi'}^{[2]})/J_A^{[1,2]}+ 
I_B^{[2,0]} (\hat{\Upsilon}_{\xi}^{[0]}, \hat{\Upsilon}_{\xi'}^{[0]})/J_B^{[2,0]}$ 
obtained in the $\eta$-$\xi$ frame has the same functional form of the mode functions and correlators as 
$I_A^{[1,0]}(\hat{\Upsilon}_x^{[0]}, \hat{\Upsilon}_y^{[0]})/J_A^{[1,0]} +  I_B^{[2,1]} 
(\hat{\Upsilon}_x^{[1]}, \hat{\Upsilon}_y^{[1]})/J_B^{[2,1]}$ in the $t$-$x$ frame. 
Thus the time order of the spacelike separated measurement events does not matter.
The wavefunctionals collapsed in two different frames are identical at $t_3=\eta_3=\pi$ up to a coordinate transformation. 

If the two measurement events are timelike separated, however, different observers in different frames will recognize 
the same time order of the two events. In this case the retarded mutual influences will come into play, so the resulting PMS 
will be different from those in the reversed time order, while the consistency of quantum states at the final time-slice 
is obvious when one knows the results for the previous cases.

\section{Summary and Discussion}

In all of the above simple cases, we found that the quantum states of the RSG detector-field system started with the same initial 
state defined on the same fiducial time-slice, then collapsed on different time-slices in different reference frames
by one or two measurements on the pointlike detectors, perhaps in different time-orders provided the measurement events 
are spacelike separated, will evolve to the same state of the combined system up to a coordinate transformation when compared 
on the same final time-slice. 

Our analysis is actually independent of any specific choice of coordinates for the spacetime between the initial and 
final time-slices or any specific property of the field theories in (1+1)D such as conformal 
symmetry\footnote{In higher dimensional spacetimes, the calculations are similar except that
the Green's functions of the field could be singular in the coincidence limit,
so when a pointlike detector is coupled to the field, one has to introduce some cutoffs and truncations to regularize the 
divergences, and some parameters of the {\it detectors} will have to be renormalized \cite{LH06}.},
and the identities interpreted as the Huygens principle for the mode functions such as $(\ref{id1})$ are consequences of 
the linearity of quantum theory and independent of the details of the initial state. 
Thus our calculations for the Gaussian states of the combined system with one or two 
detectors collapsed by one or two spatially local measurements in (1+1)D Minkowski space can be generalized straightforwardly 
to the cases with more detectors, more measurement events, and 
more general states, in higher dimensions. 

While we took advantage of the linearity of our model to perform explicit calculations, 
the consistency observed here is actually more generic in RQFT. In \ref{NLinDFT} a similar argument has been 
applied to the detector-field systems with nonlinear, spatially local interactions. One can see that the 
multi-point correlators of the combined system evaluated on the final time-slice are still independent of the data on 
the time-slice outside the detector at the moment of the measurement on the detector. 
So the quantum states in these nonlinear local detector-field theories started with the same initial state 
and then collapsed by the same spatially local projective measurements in different reference frames 
will be consistent with each other when compared on the same final time-slice.
The spatial locality of interactions (which can be linear or nonlinear) and the general covariance 
in a relativistic system, together with the linearity of quantum dynamics and 
the spatial locality of quantum measurements in its quantum theory, 
guarantee this consistency.\\ 

\noindent {\bf Acknowledgment}
The author thanks Bei-Lok Hu for illuminating discussions and Lajos Di\'osi for helpful comments. 
This work was supported by the NSC Taiwan under the grant 99-2112-M-018-001-MY3 and in part by the National Center 
for Theoretical Sciences, Taiwan.

\begin{appendix}

\section{Comparison with the indirect measurement scheme}

In the indirect measurement scheme for relativistic quantum systems \cite{AA81, BP02, Dio91}, there is no projective measurement
between the initial and final time-slices. Instead, a measurement event is modeled by linearly coupling,
and so entangling, a quantum probe or device locally in space and time with the field to be observed. 
Each local coupling process is described by an interaction Hamiltonian with a compact support 
localized in spacetime between the initial and final time-slices. 
After all these couplings or interactions have been finished, a set of the projective measurements on the quantum probes 
will be performed on the final time-slice associated with some moment essentially in the asymptotic region.

In contrast, the RSG detectors and the field in our model are interacting continuously at all times, and the projective measurements 
on the detectors, if any, will be performed at some moments between the initial and final time-slices, which will always be in the 
interaction region. Then we compare the quantum states of the combined system in different frames on the same final time-slice, 
still in the interaction region, without any further projective measurement there.

\subsection{Reference frames and quantum states without projective measurement}

In \cite{AA84} Aharonov and Albert concluded that there is no covariant description of measurement in terms of time
evolution of quantum states when the whole time-slice which the quantum state is defined on is not in the asymptotic region 
where all interactions have been finished. 
In the simplest case that they considered with no wavefunction collapse between the initial and the final 
time-slices, they have found that quantum states have to be parameterized by many local times if the system consists of many local 
quantum objects. Di\'osi pointed out that in this case it would be simpler to describe such processes using Heisenberg 
operators, which are naturally covariant in a relativistic model \cite{Dio91}.  
In our model $(\ref{RSGact})$ and other linear detector theories, this can be understood as follows. 

A quantum state in the Schr\"odinger picture for our model is represented as a wavefunctional consisting of the dynamical 
variables ($Q_{\bf d}$, $P_{\bf d}$, $\Phi_x$, and $\Pi_x$, or equivalently, $K^\mu$ and $\Delta^\mu$ in $(\ref{Qstate})$ in 
the $(K, \Delta)$-representation) and the time-dependent factors. The latter can be expressed as combinations of the correlators of the 
dynamical variables (for Gaussian states, only two-point correlators are involved; see $(\ref{Qstate})$, and also see \cite{LH07} 
for the conventional Wigner functionals.) Without any projective measurement, the form of these combinations of the correlators will 
remain unchanged from the initial all the way to the final time-slices, 
while the time dependence of the factors is due to the evolving correlators of the corresponding time-dependent, covariant operators 
with respect to the initial state of the combined system in the Heisenberg picture. 
Here the time evolution of those covariant operators (or equivalently, the mode functions in Section 
\ref{RSGdetector} in our linear system) are governed by the Heisenberg equations of motion, which are also covariant in a relativistic 
system with a well defined Hamiltonian. The evolutions of the operators corresponding to the dynamical variables local in position space, 
such as $Q_{\bf d}$ and $P_{\bf d}$ living inside 
the pointlike detectors, are naturally parameterized by their own proper times, like those in $(\ref{HOeom})$. 

Suppose the time evolutions of all the operators have been solved. In a given reference frame started with some initial time-slice, 
the quantum state on a time-slice associated with some moment in this frame is obtained by inserting the operators for the 
dynamical variables on that time-slice into the correlators composing the time-dependent factors in the wavefunctional. 
For the dynamical variables local in position space ($Q_{\bf d}$, $P_{\bf d}$, $\Phi_{\bf x}$, and $\Pi_{\bf x}$) 
those operators are the ones with their own proper times or coordinate times at the intersections of their world lines and the 
time-slice that the correlators are evaluated on. Different frames have different time-slicing schemes, so the histories of the
wavefunctionals in different frames can be very different, though the history of each operator corresponding to a spatially 
local dynamical variable is uniquely determined by the covariant Heisenberg equations of motion \cite{LCH08}. 

Note that the above argument is not restricted to just the indirect measurement scheme, but is valid for all relativistic
systems with well defined Hamiltonians without any projective measurement between the initial and final time-slices.  
The only thing special to the indirect measurement scheme is that all the couplings will eventually be turned off, 
then a set of the projective measurements will be performed on the quantum probes or devices on some final time-slice 
in the future asymptotic region. It is obvious that, if the covariant interaction Hamiltonians describing two individual local 
``measurement events" (entanglement processes, actually) in the indirect measurement scheme commute, 
then the time order of these events being possibly different in different frames will not matter and the quantum states 
will be consistent when compared on {\it the same} final time-slice in different frames in the future asymptotic region 
\cite{AA81, BP02}. This is guaranteed by the covariance of the operator histories 
and the form-invariance of the time-dependent factors in terms of the correlators in a quantum state never being projected. 

If there is a projective measurement performed at some moment during the evolution of a detector-field system in the interaction region, 
however, the dependence on the correlators of the time-dependent factors in the wavefunctional of the whole detector-field system 
will be suddenly changed like $(\ref{cort1t0})$, so the argument in this subsection will not be valid.

\subsection{Joint probability and reduced state of the detectors}

Suppose each quantum device in the indirect measurement scheme is associated with a pointlike object in space 
so that the dynamics of the device can be parameterized in its proper time, which is invariant under coordinate transformations. 
Following   \cite{Dio91}, suppose the device-field system is started with the initial state $\left|D_1\right> \otimes 
\left|D_2\right> \otimes \cdots \left|D_N\right> \otimes \left| \psi_{in}\right> $ defined on the initial time-slice in some 
reference frame, where $\left|D_n\right>$ is a Gaussian state of the $n$-th device and $\left| \psi_{in}\right>$ is the initial 
state of the field. Right after all the 
local device-field interactions are finished, say, at $t_1$ in this frame, the device-field system 
evolves into an entangled state of the devices and the field, denoted by $\left| D'_1, D'_2, \ldots, D'_N, \psi_{out}; t_1\right>$.
While there is no device-field or device-device coupling after $t_1$,
the backreactions from the devices to the field will keep evolving in the field sector,
and so do the quantum state of the combined system as well as the correlations between the devices and the field. 
After some time a set of the local projective measurements on all the devices are performed simultaneously on some final 
time-slice associated with, say, $t_2$ in the same reference frame ($t_2>t_1$) with the outcomes $\{{\bar p}^f_1, {\bar p}^f_2,\ldots, 
{\bar p}^f_N \}$, then the post-measurement state of the combined system becomes $\left|\right.{\bar p}^f_1\rangle \otimes 
\left|\right.{\bar p}^f_2\rangle \otimes \cdots \left|\right.{\bar p}^f_N\rangle \otimes {\cal N} \left| \psi_{final}\right>$
with the normalization factor ${\cal N}$ and the un-normalized final state of the field $\left| \psi_{final}\right> = \left(\prod_n 
\langle {\bar p}^f_n \left.\right| \right) \left| D'_1, D'_2, \ldots, D'_N, \psi_{out}; t_2\right>$ defined on the time-slice 
associated with $t_2$. 
The joint probability of the set of the outcomes $\{{\bar p}^f_1, {\bar p}^f_2,\ldots, {\bar p}^f_N \}$ is 
proportional to the squared norm of $\left| \psi_{final}\right>$, namely,  
\begin{eqnarray}
  w({\bar p}^f_1, {\bar p}^f_2,\ldots, {\bar p}^f_N ) &\propto& \langle \psi_{final} | \psi_{final}\rangle = 
    {\rm Tr}_{\Phi_{\bf x}} | \psi_{final}\rangle  \langle \psi_{final}| \nonumber\\
    &=&  \left(\prod_n \langle {\bar p}^f_n |\right)\rho_D^{out}(t_2)
    \left(\prod_{n'} | {\bar p}^f_{n'} \rangle\right).
\end{eqnarray}
where the reduced state of the devices is given by
\begin{equation}    
    \rho_D^{out}(t_2) \equiv {\rm Tr}_{\Phi_{\bf x}} | D'_1, D'_2, \ldots, D'_N, \psi_{out}; t_2\rangle  
    \langle D'_1, D'_2, \ldots, D'_N, \psi_{out}; t_2 | .
\end{equation}
Thus the joint probability $w({\bar p}^f_1, {\bar p}^f_2,\ldots, {\bar p}^f_N )$ taken at $t_2$ corresponds to the 
reduced state of the devices right before the projective measurements, $\rho_D^{out}(t_2)$, where the field operators defined on 
the time-slice associated with $t_2$ are traced out (or coarse-grained). By measuring the devices in an ensemble of many copies 
of the state $| D'_1, D'_2, \ldots, D'_N, \psi_{out}; t_2\rangle$, one could reconstruct the reduced state $\rho_D^{out}(t_2)$ 
of the device, rather than the quantum state of the field or the combined system, using the outcomes of the measurements. 
Here the choice of the final time-slice actually specifies a class of reference frames with all kinds of the final time-slices  
intersecting the same set of the projective measurement events localized in spacetime. The reduced states of the devices  
$\rho_D^{out}$ right before the projective measurements in all the reference frames in this particular class are the same.

Similarly, since the harmonic oscillators in the pointlike RSG detectors in our model are scalars 
parameterized in their proper times, the outcomes of the projective measurements on the detectors are automatically invariant 
under coordinate transformations. If the projective measurement events on the detectors are spacelike distributed on a time-slice  
in some reference frame, the outcomes of the measurements on an ensemble of many copies of the quantum state of the combined 
system will correspond to the joint probability as well as the reduced state of the detectors defined on that time-slice.
This reduced state of the detectors is identical to all those observed in a class of reference frames
in which each frame has a time-slice simultaneously intersecting all the projective measurement events. Indeed, 
in our model, the Gaussian reduced state of the detectors at some moment $x^0>0$ in some reference frame without any 
projective measurement between $t=\eta=0$ and $x^0$ reads
\begin{eqnarray}
    &&\rho^{}_D[K^{\bf d}, \Delta^{\bf d}; x^0] = \exp \left\{ -{1\over 2\hbar^2}  \times\right. \nonumber\\ &&\left.
    \sum_{\bf d,d'}\left[ K^{\bf d} {\cal Q}_{\bf d d'}(x^0) K^{\bf d'} 
    -2 \Delta^{\bf d} {\cal R}_{\bf d d'}(x^0) K^{\bf d'} + 
    \Delta^{\bf d} {\cal P}_{\bf d d'}(x^0) \Delta^{\bf d'} \right]\right\},
\end{eqnarray}
where ${\bf d}$ and ${\bf d'}$ run over all the detectors, and the factors ${\cal Q}_{\bf d d'}$, ${\cal R}_{\bf d d'}$, and 
${\cal P}_{\bf d d'}$ are the symmetric two-point correlators of the detectors Tr$[\{R^{}_{\bf d}(\tau^{}_{\bf d}(x^0)),$
$R^{}_{\bf d'}(\tau^{}_{\bf d'}(x^0))\}\rho(0)]$ with $R=Q, P$ 
(cf. (\ref{Qstate})). From $(\ref{Qexp})$, $(\ref{zerophi})$ and $(\ref{eomHO1})$ one can see that only the operators in the past lightcones 
of the measurement events are relevant in $\rho^{}_D[K^{\bf d}, \Delta^{\bf d}; x^0]$, so the reduced state of the detectors 
is independent of the choice of the time-slice outside the measurement events. 

In contrast to the detector sector,
the field sector of the collapsed or projected state of the combined system (e.g. the $\left| \psi_{final}\right>$ above)
defined on different time-slices passing through the same set of events of the projective measurements 
on the devices or the detectors
can appear very differently but cannot be compared directly. Nevertheless, our result in this paper implies that if these 
collapsed states can be compared later on some time-slice in common in those different frames, they will be identical to 
each other up to a coordinate transformation. 

Note that in the position space representation of the field, each field amplitude defined at a space point corresponds to one 
degree of freedom, and there are infinitely many degrees of freedom for the field, as many as the number of the space points
in the universe. It is impossible to obtain the full 
information of the field through finitely many measurement events on a finite number of the devices or the detectors.
So our result in this paper on the consistency of the quantum state of the whole detector-field system
is certainly stronger then those on the reduced state of the detectors.

\section{Local measurement in nonlinear detector-field theories}
\label{NLinDFT}

A similar argument can be applied to the nonlinear detector-field theories where
all the detector-field and field-field couplings are spatially local. 
Here we illustrate the single-detector case with one measurement on the detector. 
The more general cases are straightforward. 

Let $(\hat{\Theta}_A, \hat{\Theta}_{A^*}, \hat{\Theta}_{x}, \hat{\Theta}_{x^*} )\equiv (\hat{Q}_A, \hat{P}_A, \hat{\Phi}_x, \hat{\Pi}_x)$
be the covariant operators in a nonlinear relativistic local detector-field theory.
Given the Hamiltonian $\hat{H}$ 
defined at the initial moment $t=t_0$ in the Minkowski frame, 
the time evolution of the operator $\hat{\Theta}_\mu$ from $t_0$ to some $t_1 > t_0$ can be expressed as
\begin{equation}
  \hat{\Theta}_\mu^{[10]} 
  =\hat{\Theta}_\mu^{[0]} + {i\over\hbar}(t_1-t_0)[\hat{H}^{[0]}, \hat{\Theta}_\mu^{[0]}] + 
  {1\over 2}\left({i\over\hbar}(t_1-t_0)\right)^2 [\hat{H}^{[0]}, [\hat{H}^{[0]}, \hat{\Theta}_\mu^{[0]}]]
  + \cdots ,
\label{Psi10}
\end{equation}
or
\begin{equation}
   \hat{\Theta}_\mu^{[10]} \equiv
    \varphi_\mu^{\nu[10]}\hat{\Theta}_\nu^{[0]} +\varphi_\mu^{\nu\sigma [10]}\{ \hat{\Theta}_{\nu}^{[0]}, \hat{\Theta}_{\sigma}^{[0]} \}
   + \varphi_\mu^{\nu\sigma\rho [10]}\{ \hat{\Theta}_{\nu}^{[0]}, \hat{\Theta}_{\sigma}^{[0]}, \hat{\Theta}_{\rho}^{[0]}\} + \cdots .
\label{varphiMF}
\end{equation}
If the operators keep evolving from $t_1$ to some $t_2 > t_1$, then one also has
\begin{eqnarray}
  \hat{\Theta}_\mu^{[20]} = \varphi_\mu^{\nu[21]}\hat{\Theta}_\nu^{[10]} +
  \varphi_\mu^{\nu\sigma [21]}\{ \hat{\Theta}_{\nu}^{[10]}, \hat{\Theta}_{\sigma}^{[10]} \}
   + \varphi_\mu^{\nu\sigma\rho [21]}\{ \hat{\Theta}_{\nu}^{[10]}, \hat{\Theta}_{\sigma}^{[10]}, \hat{\Theta}_{\rho}^{[10]}\} + \cdots,
\label{Psi20}
\end{eqnarray}
which is a consequence of the linearity of quantum theory.
Here the brace bracket $\{\hat{\Theta}_1, \hat{\Theta}_2, \cdots\}$ denotes the fully symmetrized ordering of the operators 
$\hat{\Theta}_1, \hat{\Theta}_2,\cdots$, the indices $\mu, \nu, \rho, \sigma = A, A^*, \{x\}, \{x^*\}$, and 
DeWitt notation is understood. The c-number factors $\varphi_\mu^\nu(t)$ are the counterparts of the mode functions in 
(\ref{Qexp}) and (\ref{Phiexp}) in the linear RSG detector theory. 

A general quantum state of the combined system at the moment $t$ in the $(K,\Delta)$-representation can be written as
\begin{equation}
  \rho({\cal K} ;t) = \exp \left( {\cal S}_\mu(t) {\cal K}^\mu + {1\over 2!}{\cal S}_{\mu\nu}(t) {\cal K}^\mu {\cal K}^\nu
   + {1\over 3!} {\cal S}_{\mu\nu\sigma}(t) {\cal K}^\mu {\cal K}^\nu {\cal K}^\sigma + \cdots  \right),
  \label{GQS}
\end{equation} 
where $( {\cal K}^A,{\cal K}^{A^*},{\cal K}^x,{\cal K}^{x^*}) \equiv ( {i\over\hbar} K^A , -{i\over\hbar}\Delta^A,
{i\over\hbar}K^x, -{i\over\hbar}\Delta^x)$.
At $t=t_1$, a measurement on detector $A$ projects the quantum state from $\rho({\cal K}, t_1)$ to the PMS
\begin{equation}
  \tilde{\rho}({\cal K}) = \rho^{}_A({\cal K}^a) N \int {d\tilde{\cal K}^A d\tilde{\cal K}^{A^*}\over 2\pi\hbar}
  \rho_A^*(\tilde{\cal K}^a) 
  \rho(\tilde{\cal K}^a, {\cal K}^X;t_1),
\end{equation}
where $N$ is the normalization factor,  $a = A$ or $A^*$, and $X \in \{x\}\cup\{x^*\}$. 
Starting with this new initial state, the combined system keeps evolving to the final time-slice associated with 
$t=t_2$, where it can be compared with the quantum states evolving in other reference frames. 
At $t_2$ the quantum state can again be written in the general form (\ref{GQS}),
where the factors ${\cal S}^{[2]}_{\mu_1,\cdots,\mu_n}\equiv {\cal S}_{\mu_1,\cdots,\mu_n}(t_2)$ in $\rho({\cal K} ;t_2)$ 
are functions of the $1$-, $2$-, $\cdots$, $n$-point symmetric correlators for the PMS right after $t_1$
such as 
\begin{eqnarray}
 &&\langle \{ \hat{\Theta}_{\mu^{}_1}^{[21]}, \hat{\Theta}_{\mu^{}_2}^{[21]},\cdots,\hat{\Theta}_{\mu^{}_n}^{[21]} \}\rangle^{}_1 
 \nonumber\\ &=& \left< \left\{ \left[\varphi_{\mu^{}_1}^{\nu[21]}\hat{\Theta}_{\nu}^{[1]} +
   \varphi_{\mu^{}_1}^{\nu\sigma [21]} \{\hat{\Theta}_{\nu}^{[1]}, \hat{\Theta}_{\sigma}^{[1]}\}+ \cdots\right],\cdots
   \right.\right.\nonumber\\ & & \,\,\, \left.\left. \cdots , 
   \left[\varphi_{\mu^{}_n}^{\nu'[21]}\hat{\Theta}_{\nu'}^{[1]} +
   \varphi_{\mu^{}_n}^{\nu'\sigma' [21]} \{\hat{\Theta}_{\nu'}^{[1]}, \hat{\Theta}_{\sigma'}^{[1]}\}
   + \cdots\right]\right\}\right>^{}_1 \nonumber\\
   &=& \left\{ \left(\varphi_{\mu^{}_1}^{\nu[21]}{\delta\over\delta{\cal K}^\nu} +\varphi_{\mu^{}_1}^{\nu\sigma [21]}
     {\delta\over\delta{\cal K}^\nu}{\delta\over\delta{\cal K}^\sigma} + \cdots\right)\cdots \right.\nonumber\\
     & & \,\,\, \left.\cdots  \left(\varphi_{\mu^{}_n}^{\nu'[21]}{\delta\over\delta{\cal K}^{\nu'}}
     +\varphi_{\mu^{}_n}^{\nu'\sigma' [21]}{\delta\over\delta{\cal K}^{\nu'}}{\delta\over\delta{\cal K}^{\sigma'}} + 
     \cdots\right)\tilde{\rho}({\cal K}) \right\}_{{\cal K}={\bf 0}}\nonumber\\ 
   &=& N\int{d\tilde{\cal K}^A d\tilde{\cal K}^{A^*} \over 2\pi\hbar} 
   \rho_A^*(\tilde{\cal K}^a)\cdot \rho({\cal K}^X=0,\tilde{\cal K}^a; t_1 )
   \left\{ \left[\langle \hat{\Upsilon}^{}_{\mu^{}_1:\mu^{}_n}\rangle^{}_0 +\right.\right.\nonumber\\ & & 
   \tilde{\cal K}^a \left(\langle \hat{\Upsilon}^{}_{\mu^{}_1:\mu^{}_n}, \hat{\Theta}_a^{[10]} \rangle^{}_0 
     -\langle \hat{\Upsilon}^{}_{\mu^{}_1:\mu^{}_n}\rangle^{}_0 \langle\hat{\Theta}_a^{[10]}\rangle^{}_0\right)+\nonumber\\  & &
   \tilde{\cal K}^a \tilde{\cal K}^b \left(
   \langle \hat{\Upsilon}^{}_{\mu^{}_1:\mu^{}_n}, \hat{\Theta}_a^{[10]}, \hat{\Theta}_b^{[10]} \rangle^{}_0 
   -\langle \hat{\Upsilon}^{}_{\mu^{}_1:\mu^{}_n}\rangle^{}_0 \langle\hat{\Theta}_a^{[10]}, \hat{\Theta}_b^{[10]} \rangle^{}_0 \right.
   \nonumber\\ & & \hspace{.5cm} \left.
   -2 \langle \hat{\Upsilon}^{}_{\mu^{}_1:\mu^{}_n}\hat{\Theta}_a^{[10]}\rangle^{}_0 \langle \hat{\Theta}_b^{[10]} \rangle^{}_0 
   +2 \langle \hat{\Upsilon}^{}_{\mu^{}_1:\mu^{}_n}\rangle^{}_0 \langle \hat{\Theta}_a^{[10]}\rangle^{}_0 \langle \hat{\Theta}_b^{[10]} \rangle^{}_0
   \right) +\nonumber\\   & & \left.\left. \cdots \right]
   \rho^{}_A({\cal K}^a)\right\}_{{\cal K}^a=0}
\label{SerRep}
\end{eqnarray}
with $\hat{\Upsilon}_{\mu^{}_1:\mu^{}_n}\equiv \{ \hat{\Upsilon}_{\mu^{}_1}, \hat{\Upsilon}_{\mu^{}_2},\cdots,
\hat{\Upsilon}_{\mu^{}_n}\}$, where
\begin{equation}
  \hat{\Upsilon}_{\mu} \equiv \varphi_\mu^{\nu[21]}\tilde{\Theta}_\nu + 
  \varphi_\mu^{\nu\sigma [21]}\{ \tilde{\Theta}_{\nu}, \tilde{\Theta}_{\sigma}\}
   + \varphi_\mu^{\nu\sigma\rho [21]}\{ \tilde{\Theta}_{\nu}, \tilde{\Theta}_{\sigma}, \tilde{\Theta}_{\rho} \} + \cdots,
\label{up}
\end{equation}
which looks similar to $\hat{\Theta}_\mu^{[20]}$ in (\ref{Psi20}) but now $\tilde{\Theta}^{}_X \equiv \hat{\Theta}_X^{[10]}$
and $\tilde{\Theta}_a \equiv {\delta\over\delta{\cal K}^a}$. From (\ref{Psi20}) and (\ref{up}), one has
\begin{eqnarray}
    & &\hat{\Upsilon}_{\mu} -\hat{\Theta}_\mu^{[20]} = 
    \varphi_\mu^{a [21]} \left( {\delta\over\delta{\cal K}^a}-\hat{\Theta}_a^{[10]}\right)
    - 2 \varphi_\mu^{aX [21]}\{ \hat{\Theta}_{a}^{[10]}, \hat{\Theta}_{X}^{[10]}\} \nonumber\\ &&\,\,\,\,+ 
    \varphi_\mu^{ab [21]} \left({\delta\over\delta{\cal K}^a}{\delta\over\delta{\cal K}^b}
    + 2 \hat{\Theta}_a^{[10]} {\delta\over\delta{\cal K}^b}-\{ \hat{\Theta}_{a}^{[10]}, \hat{\Theta}_{b}^{[10]}\}\right)
    + \cdots.
\label{upDef}
\end{eqnarray}
Comparing (\ref{Psi10}) with (\ref{varphiMF}), one can see that the factors
$\varphi_\mu^{a_1\cdots a_m X_1\cdots X_n [21]}$ with $m, n \ge 1$ vanish whenever $X_j \not= z_A^1(t_1)$ or 
$z_A^{1*}(t_1)$, $j=1,2,\cdots,n$, because the detector-field coupling is spatially local. 
This implies that no operator defined on the $t_1$-slice outside of detector $A$ 
is present on the right hand side of (\ref{upDef}).
Since $\hat{\Theta}_\mu^{[20]}$ is independent of the $t_1$-slice, 
$\hat{\Upsilon}_{\mu}$ is independent of the operators on the $t_1$-slice outside of detector $A$.
Thus $\langle \{ \hat{\Theta}_{\mu^{}_1}^{[21]},\hat{\Theta}_{\mu^{}_2}^{[21]},\cdots,
\hat{\Theta}_{\mu^{}_n}^{[21]}\} \rangle^{}_1$ 
are independent of the data on the $t_1$-slice outside of detector $A$,
and so is the quantum state $\rho({\cal K}; t_2)$ of the combined system.

\end{appendix}


\end{document}